\documentclass[11pt,twoside]{article}

\usepackage{asp2006}
\usepackage{graphicx}
\usepackage{epsf}
\usepackage{psfig}
\usepackage{lscape}

\begin{document}
\title{Concluding Remarks}   %%% Fill in title
\author{Slavek M. Rucinski}                           %%% Fill in author names
\affil{Department of Astronomy and Astrophysics,
University of Toronto, Toronto, ON, M5S~3H4, Canada}  %%% Fill in author affiliations

\begin{abstract}                                    %%% Abstract to run on from here.
Version 3. There should be no abstract, I think.
\end{abstract}

%\section{}   %%% Top level section head (remove "%" symbol)
%\subsection{}   %%% Second level section head (remove "%" symbol)
%\subsubsection{}   %%% Lowest level section head (remove "%" symbol)
%\section*{}    %%% Unnumbered top level section head (remove "%" symbol)
%\subsection*{}   %%% Unnumbered second level section head (remove "%" symbol)

This was an useful conference. It has shown us the rapidly widening scope of
the binary star research which now extends in its applications from the
realm of planets to black holes and binary galactic nuclei. The binary
star domain appears to be -- in parallel to searched for extra-solar
planets and star formation -- the most active areas of stellar
astrophysics. I am glad I attended this conference.

When the Chairman of the Scientific Organizing Committee
asked me give the final conclusions, I felt puzzled and surprised. I am
not used to giving summary talks and never considered myself
being able to deliver a reasonable set of conclusions after any
conference I attended; I had a lot of respect for people who
could do that. After much deliberation,
I came up with three serious questions for myself:
(1)~What and how to summarize? (2)~Do I really have anything useful to say?
(3)~Should I be nice or honest?

Answering the first question, I realized that
I see no way to formulate a summary or overview of what
was discussed: That would be impossible to
achieve for such a broad and rapidly expanding field. There were
too many talks and I would do a grave disservice if I tried to single
out those that I liked and then ignore those that I did not notice.
A ready solution would be then to say a few platitudes and
every body would be happy. But would this be useful?
Instead, I decided to be a bit provocative and perhaps even
unpleasant by saying things we normally do not want to hear,
especially at the end of a successful conference.
I am old enough to have a comfort of saying what I think...
You may consider my conclusions or rather observations
to read when having some time to kill; in any case, they
are simple, if not obvious:
\begin{itemize}
\item I see a clear danger for the binary-star
community of becoming irrelevant by defining
too narrow horizons and too particular goals to ourselves.
We must honest about that:
Binary stars are no longer an attractive area of study: This
currently belongs to cosmology and extra-solar planets.
We should stay close to both fields.
%There may be other, broader and more
%universally relevant applications where out expertise may
%be relevant, but our community would have to work hard
%to overcome its weaknesses and fragmentation.
\item Modeling of the data must have a reason. We are used to
a mode of operation that observers collect
data and then interpret them using models.
I fear that sometimes the models are too
mechanically applied. We should always think about the use
of models as a method to simplify -- i.e.\ understand -- of
what is going on. Too complex or too routine modeling
make our understanding harder and confused.
\item The binary star community must stay focused. Are we focused
enough? Do we know what we want to achieve? Or are we observing
and studying binary stars by some sort of inertia, ``because
they are there''? This may be not enough in the future and
we must think about a multitude of new branches of astronomy
competing for finding and taxpayers acceptance.
\item Are we really ready for the flood of data coming from
new surveys and new satellites? I fear that the current
binary star community may one day find itself excluded from
such data sources: New people will come and use these data
before we will realize that...
\item One of the most depressing views to me
is to see papers submitted (and published!) about single,
basically uninteresting eclipsing binaries analyzed using
the same, monotonously repeated
approaches. Usually, such ``cookie-cutter'' papers contain
the same, boring and repetitive litany describing which
synthesis model and which darkening (gravity, limb) darkening
coefficients were used. Such papers do not achieve much
except padding of publication lists: Nobody 
%(except some good-natured but misled theoreticians) 
will ever use these ``solutions''.
Why not to prepare for large-volume studies by analyzing
many binaries in one go? The ASAS, OGLE, MACHO surveys
with tens of thousands of light curves
are an excellent practice ground for us before the flood of
millions of light curves will appear at our doorsteps.
\item And last -- but not least -- concerning this
 conference: Many presentations were poor, read form the screen,
with no eye-contact, frequently leaving an impression of being
prepared in a haste. In many cases
the number of slides by far exceeded the highest
imaginable delivery rate even in speaker's native
tongue. How can one ever dream of presenting
50 slides in 20 minutes (including a discussion!) as it was
attempted during the conference?
My advice is: Take the matter seriously and prepare well.
Time your presentations in advance
(at least 2 minutes per slide or slower);
memorize it if the English is not your mother tongue;
do not over-estimate your abilities to speak publicly,
especially if you do not lecture on a routine basis.
And -- frankly -- do not give a presentation if you
cannot do it well.
% because you may badly damage your reputation.
After all, a good poster is much
better than a poor oral presentation.
\end{itemize}

It was good to see old friends and to make new acquaintances.
Young active astronomers are ready to replace the ``old
guard'' and this is great. Sorry for the minor tone
of my remarks; perhaps they are misplaced.
But we should never feel too satisfied with ourselves...

The Scientific Organizing Committee, led by Dr.\ Andrej
Pr\v{s}a, and the Local Organizing
Committee, led by Dr.\ Miloslav Zejda,
prepared a memorable meeting. Many thanks to them.
My particular respect is reserved to the LOC for their
quiet, low-key efficiency.  A lot of
preparation went into organizing this meeting yet their
work seemed effortless and rather invisible. One of
the signs of the excellent preparation for the
meeting was the ease of hooking up various
computers to the display system: It felt so simple as if
this has not been one the most common sources of time loss
in lecture halls and at conferences.
Thank you very much.

%\acknowledgements %%% Text of acknowledgements runs on after this command.

%%% THE BIBLIOGRAPHY
%%%
%%% CONSULT SECTION 3 OF "INSTRUCTIONS FOR AUTHORS" FOR HOW TO USE NATBIB.
%%% AUTHORS ARE ENCOURAGED TO USE EITHER THE "THEBIBLIOGRAPY" ENVIRONMENT
%%% BY UNCOMMENTING (DELETING THE "%" SYMBOL) THE COMMANDS BELOW, OR BY
%%% USING THE BIBTEX ENVIRONMENT. TO FIND OUT WHICH IS APPLICABLE TO YOUR
%%% CONTRIBUTION, CONSULT THE VOLUME EDITORS FOR YOUR PROCEEDINGS.
%%%

%\begin{thebibliography}{}
%\bibitem[]{}
%\bibitem[]{}
%\bibitem[]{}
%\bibitem[]{}
%\bibitem[]{}
%\bibitem[]{}
%\bibitem[]{}
%\bibitem[]{}
%\bibitem[]{}
%\bibitem[]{}
%\bibitem[]{}
%\bibitem[]{}
%\end{thebibliography}

\end{document}